\documentclass[12pt,preprint]{aastex}

\newcommand{\wfcfuv}{$U_{\rm 225}$}
\newcommand{\wfcnuv}{$U_{\rm 275}$}
\newcommand{\wfcuv}{$U_{\rm 336}$}
\newcommand{\acsb}{$B_{\rm 435}$}
\newcommand{\acsv}{$V_{\rm 606}$}
\newcommand{\acsi}{$i_{\rm 775}$}
\newcommand{\acsz}{$z_{\rm 850}$}
\newcommand{\wfcy}{$Y_{\rm 098}$}
\newcommand{\wfcj}{$J_{\rm 125}$}
\newcommand{\wfch}{$H_{\rm 160}$}
\newcommand{\sqmin}{arcmin$^2$}
\newcommand{\uvdrops}{$z\simeq 1$--$3$}
\newcommand{\bvdrops}{$z\simeq 4$--$5$}
\newcommand{\etal}{{et\thinspace al.}}
\newcommand{\Ho}{$H_{\rm 0}$}
\newcommand{\tabref}[1]{Table~\ref{#1}}
\newcommand{\figref}[1]{Figure~\ref{#1}}
\newcommand{\secref}[1]{\S~\ref{#1}}

\begin{document}

\title{Stellar Populations of Lyman Break Galaxies at
  \uvdrops\ in the HST/WFC3
  Early Release Science Observations}

\shorttitle{Stellar Populations of the UV-dropouts}

\author{
N.~P.~Hathi\altaffilmark{1},  
S.~H.~Cohen\altaffilmark{2},
R.~E.~Ryan~Jr.\altaffilmark{3}, 
S.~L.~Finkelstein\altaffilmark{4},
P.~J.~McCarthy\altaffilmark{1},
R.~A.~Windhorst\altaffilmark{2},
H.~Yan\altaffilmark{5}, 
A.~M.~Koekemoer\altaffilmark{3},
M.~J.~Rutkowski\altaffilmark{2},
R.~W.~O'Connell\altaffilmark{6},
A.~N.~Straughn\altaffilmark{7},
B.~Balick\altaffilmark{8},
H.~E.~Bond\altaffilmark{3},
D.~Calzetti\altaffilmark{9},
M.~J.~Disney\altaffilmark{10},
M.~A.~Dopita\altaffilmark{11,12},
Jay~A.~Frogel\altaffilmark{12,13},
D.~N.~B.~Hall\altaffilmark{14},
J.~A.~Holtzman\altaffilmark{15},
R.~A.~Kimble\altaffilmark{7},
F.~Paresce\altaffilmark{16},
A.~Saha\altaffilmark{17},
J.~I.~Silk\altaffilmark{18},
J.~T.~Trauger\altaffilmark{19},
A.~R.~Walker\altaffilmark{20},
B.~C.~Whitmore\altaffilmark{3},
and
E.~T.~Young\altaffilmark{21}
}

\altaffiltext{1}{Observatories of the Carnegie Institution for Science,
  Pasadena, CA 91101, USA}

\altaffiltext{2}{School of Earth and Space Exploration, Arizona State
  University, Tempe, AZ 85287-1404, USA}

\altaffiltext{3}{Space Telescope Science Institute, Baltimore, MD
  21218, USA}

\altaffiltext{4}{Department of Astronomy, The University of Texas,
  Austin, TX 78712, USA}

\altaffiltext{5}{Department of Physics \& Astronomy, University of
  Missouri, Columbia, MO 65211, USA}

\altaffiltext{6}{Department of Astronomy, University of Virginia,
Charlottesville, VA 22904-4325, USA}

\altaffiltext{7}{NASA--Goddard Space Flight Center, Greenbelt, MD
  20771, USA}

\altaffiltext{8}{Department of Astronomy, University of Washington,
Seattle, WA 98195-1580, USA}

\altaffiltext{9}{Department of Astronomy, University of Massachusetts,
Amherst, MA 01003, USA}

\altaffiltext{10}{School of Physics and Astronomy, Cardiff University,
 Cardiff CF24 3AA, UK}

\altaffiltext{11}{Research School of Astronomy \& Astrophysics,  The
Australian National University, ACT 2611, Australia}

\altaffiltext{12}{Astronomy Department, King Abdulaziz University,
  P.O. Box 80203, Jeddah, Saudi Arabia} 

\altaffiltext{13}{Galaxies Unlimited, 1 Tremblant Court,
  Lutherville, MD 21093, USA} 

\altaffiltext{14}{Institute for Astronomy, University of Hawaii,
Honolulu, HI 96822, USA}

\altaffiltext{15}{Department of Astronomy, New Mexico State
University, Las Cruces, NM 88003, USA}

\altaffiltext{16}{Istituto di Astrofisica Spaziale e Fisica Cosmica,
  INAF, Via Gobetti 101, 40129 Bologna, Italy }

\altaffiltext{17}{National Optical Astronomy Observatories, Tucson, AZ
85726-6732, USA}

\altaffiltext{18}{Department of Physics, University of Oxford, Oxford
OX1 3PU, UK}

\altaffiltext{19}{NASA--Jet Propulsion Laboratory, Pasadena, CA 91109,
USA}

\altaffiltext{20}{Cerro Tololo Inter-American Observatory,
La Serena, Chile}

\altaffiltext{21}{NASA--Ames Research Center, Moffett Field, CA 94035,
USA}

\email{nhathi@obs.carnegiescience.edu}
\shortauthors{Hathi et al}


\begin{abstract}

  We analyze the spectral energy distributions (SEDs) of Lyman break
  galaxies (LBGs) at \uvdrops\ selected using the \emph{Hubble Space
    Telescope} (\emph{HST}) Wide Field Camera 3 (WFC3) UVIS channel
  filters. These \emph{HST}/WFC3 observations cover about 50~\sqmin\
  in the GOODS-South field as a part of the WFC3 Early Release Science
  program. These LBGs at \uvdrops\ are selected using dropout
  selection criteria similar to high redshift LBGs. The deep
  multi-band photometry in this field is used to identify best-fit SED
  models, from which we infer the following results: (1) the
  photometric redshift estimate of these dropout selected LBGs is
  accurate to within few percent; (2) the UV spectral slope $\beta$ is
  redder than at high redshift ($z>3$), where LBGs are less dusty; (3)
  on average, LBGs at \uvdrops\ are massive, dustier and more highly
  star-forming, compared to LBGs at higher redshifts with similar
  luminosities ($0.1L^*\lesssim L \lesssim 2.5L^*$), though their
  median values are similar within 1$\sigma$ uncertainties. This could
  imply that identical dropout selection technique, at all redshifts,
  find physically similar galaxies; and (4) the stellar masses of
  these LBGs are directly proportional to their UV luminosities with a
  logarithmic slope of $\sim$0.46, and star-formation rates are
  proportional to their stellar masses with a logarithmic slope of
  $\sim$0.90. These relations hold true --- within luminosities probed
  in this study --- for LBGs from $z\simeq 1.5$ to $5$. The
  star-forming galaxies selected using other color-based techniques
  show similar correlations at $z\simeq 2$, but to avoid any selection
  biases, and for direct comparison with LBGs at $z>3$, a true Lyman
  break selection at $z\simeq 2$ is essential. The future \emph{HST}
  UV surveys, both wider and deeper, covering a large luminosity range
  are important to better understand LBG properties, and their
  evolution.

\end{abstract}



\section{Introduction}\label{introduction}

The high redshift frontier has moved to $z>7$ as a result of the high
resolution near-infrared (NIR) images from the \emph{Hubble Space
  Telescope} (\emph{HST}) Wide Field Camera 3 (WFC3), and the Lyman
break `dropout' technique.  The Lyman break technique was first
applied to select Lyman break galaxies (LBGs) at $z\simeq 3$
\citep{guha90,stei96,stei99}, and since then it has been extensively
used to select and study LBG candidates at redshifts $z\simeq 3$--$8$
\citep[e.g.,][]{bouw07,hath08b,redd09,fink10,yan10}.  This dropout
technique has generated large samples of faint star-forming galaxy
candidates at $z\simeq 3$--$8$. However, at highest redshifts ($z>3$),
it is very difficult to understand the details of their stellar
populations using current space and ground-based telescopes. Their
faint magnitudes make it extremely difficult to do spectroscopic
studies, and limited high resolution rest-frame optical photometry
make it challenging to investigate their spectral energy distributions
(SEDs).  These limitations make it imperative to identify and study
LBGs at lower redshifts ($z\lesssim 3$). The primary reason for the
lack of dropout selected LBGs at \uvdrops\ is that we need highly
sensitive space-based cameras to observe the mid- to near-ultraviolet
(UV) wavelengths required to identify Lyman break at \uvdrops.  The
peak epoch of global star-formation rate at \uvdrops\ is now
accessible using the dropout technique with the WFC3 UVIS
channel. \citet[][hereafter H10]{hath10} and \citet{oesc10} have used
the \emph{HST} WFC3 with its superior sensitivity to photometrically
identify lower redshift (\uvdrops) LBGs.  Understanding the LBGs at
$z\lesssim 3$ is vital for two main reasons.  First, we need to study
the star-formation properties of these LBGs, because they are at
redshifts corresponding to the peak epoch of the global star-formation
rate \citep[e.g.,][]{redd08,redd09,ly09,bouw10}.  Second, they are likely lower
redshift counterparts of the high redshift LBGs --- because of their
identical dropout selection and similar physical properties --- whose
understanding will help shed light on the process of reionization in
the early universe \citep[e.g.,][]{labb10,star10}.

There are primarily three techniques to select star-forming galaxies
at $z\simeq 2$: (1) \emph{sBzK} \citep[using the $B$, $z$, $K$
  bands,][]{dadd04,dadd07}, (2) BX/BM \citep[using the $U$, $G$, $R$
  bands,][]{stei04, adel04}, and (3) LBG \citep[using the bands which
  bracket the redshifted Lyman limit, H10;][]{oesc10}.  All these
approaches select star-forming galaxies, and yield insight into the
star-forming properties of these galaxies, but they have differing
selection biases, and so these samples don't completely overlap
\citep[see][for details]{ly11,habe12}. Therefore, it is essential to
apply identical selection criteria at all redshifts to properly
compare galaxy samples and accurately trace their evolution.  The LBG
selection is widely used to select high redshift ($z>3$) galaxies, and
to do equal comparison with these galaxies, here we investigate
physical properties of LBGs at $z\lesssim 3$.

H10 used UV observations of the WFC3 Science Oversight Committee 
Early Release Science extragalactic program (PID: 11359, PI:
O'Connell; hereafter ``ERS''), which covers approximately 50~\sqmin\
in the northern-most part of the Great Observatories Origins Deep
Survey \citep[GOODS;][]{giav04} South field, to identify LBGs at
\uvdrops. The high sensitivity of the WFC3 UVIS channel data
\citep{wind11}, along with existing deep optical data obtained with
the Advanced Camera for Surveys (ACS) as part of the GOODS program are
ideal to apply dropout technique in observed UV filters to select LBG
candidates at \uvdrops.  In this paper, we use this H10 sample of LBGs to
investigate their physical properties by fitting stellar synthesis
models to their observed SEDs.

This paper is organized as follows: In \secref{data}, we summarize the
WFC3 ERS observations, and discuss our LBG sample at \uvdrops\ as well
as the comparison sample of LBGs at \bvdrops. In
\secref{seds}, we fit observed SEDs of LBGs at \uvdrops\ and \bvdrops\
to stellar population synthesis models, and discuss
the best-fit parameters (redshift, UV spectral slope, stellar mass,
stellar age, and star-formation rates) obtained from these SED fits.
In \secref{results}, we discuss correlations between best-fit physical
parameters and their implications on our understanding of LBGs.
In \secref{conclusion}, we conclude with a summary of our
results.

In the remaining sections of this paper we refer to the
\emph{HST}/WFC3 F225W, F275W, F336W, F098M, F125W, F160W, filters as
\wfcfuv, \wfcnuv, \wfcuv, \wfcy, \wfcj, \wfch, to the \emph{HST}/ACS
F435W, F606W, F775W, F850LP filters as \acsb, \acsv, \acsi, \acsz, and
to the \emph{Spitzer}/IRAC 3.6~$\mu$m, 4.5~$\mu$m filters as [3.6],
[4.5], respectively, for convenience.  We assume a \emph{Wilkinson
  Microwave Anisotropy Probe} (WMAP) cosmology with $\Omega_m$=0.274,
$\Omega_{\Lambda}$=0.726 and \Ho=70.5~km s$^{-1}$ Mpc$^{-1}$, in
accord with the 5 year WMAP estimates of \citet{koma09}.  This
corresponds to a look-back time of 10.4~Gyr at $z\simeq 2$.
Magnitudes are given in the AB$_{\nu}$ system \citep{oke83}.


\section{Observations and Sample Selection}\label{data}

The WFC3 ERS observations \citep{wind11} were done in both the UVIS
(with a FOV of 7.30 \sqmin) and the IR (with a FOV of 4.65 \sqmin)
channels. Here, we briefly summarize the UV imaging observations.  The
WFC3 ERS UV observations were carried out in three broad-band filters
\wfcfuv, \wfcnuv\ and \wfcuv. The \wfcfuv\ and \wfcnuv\ filters were
observed for 2 orbits ($\sim$5688s) per pointing, while the \wfcuv\
filter was observed for 1 orbit ($\sim$2778 s) per pointing, for a
total of 40 orbits over the full ERS field (8 pointings).  We used the existing
GOODS v2.0\footnote{http://archive.stsci.edu/pub/hlsp/goods/v2/}
reduction of the ACS images in four optical bands (\acsb, \acsv,
\acsi, \acsz), which were re-binned to a pixel size of 0.09\arcsec.
To match the ERS IR (\wfcy, \wfcj, \wfch) and re-pixellated ACS
optical images, the UV mosaics have a pixel scale of
0.090\arcsec~pix$^{-1}$ and cover $\sim$50~\sqmin\ area of the
GOODS-South field.  Details of these observations and reduction
process are described in \citet{wind11}.

The combination of the three WFC3 UV filters and the four ACS optical
filters provide an excellent ability to select LBGs at \uvdrops\
\citep[H10;][]{oesc10}, using the dropout technique to detect the
Lyman-break at rest-frame 912~\AA\ \citep{mada95}. H10 used dropout
color selection technique in the ERS UV field to identify three sets
of UV-dropouts --- \wfcfuv-dropouts, \wfcnuv-dropouts and
\wfcuv-dropouts --- which are LBG candidates at $z\simeq 1.6$,
$2.2$ and $2.6$, respectively. They found 66 \wfcfuv-, 151 \wfcnuv-
and 256 \wfcuv-dropouts to a magnitude limit of AB$\,\simeq\,$26.5~mag,
respectively.

In this paper, we start with the H10 LBG sample. Our goal is to
investigate SEDs of reliable LBG candidates with at least 10-band HST
coverage (augmented by additional data as described in \secref{seds})
from the H10 sample, so we apply the following filtering
criteria. First criterion is the availability of the WFC3 IR (\wfcy,
\wfcj, \wfch) data. H10 used the WFC3 UVIS and ACS data to select LBGs
at \uvdrops. The WFC3 UVIS channel has a larger final ERS mosaic than
the WFC3 IR channel, so we exclude LBG candidates that don't have WFC3
IR data from our SED analysis. This criterion reduces the H10 sample
size by about 10\%.  Secondly, galaxies with poor SED fits (see
\secref{seds}) as measured by their larger $\chi^2$ were excluded from
the sample.  The galaxies which fail the SED fit are usually fainter,
does not have all NIR photometric data or have highly uncertain NIR
photometry, has poor best-fit SED (indicated by high $\chi^2$) and
could have a primary lower-redshift ($z<1$) solution.  This criterion
removes additional $\sim$10\% of galaxies from the H10 sample. This
fraction of catastrophic $\chi^2$ outliers is consistent with the
outlier fraction in the photometric redshift distribution of the
dropouts in the H10 sample.  The final sample of LBGs for the SED
analysis --- after applying above mentioned criteria --- is 47
\wfcfuv-, 126 \wfcnuv- and 213 \wfcuv-dropouts.

To compare SED properties of LBGs at \uvdrops, we select \acsb- and
\acsv-dropouts in the WFC3 ERS field. These dropouts --- LBG
candidates at $z\simeq 3.7$ and $4.7$, respectively --- were selected
following the \citet{bouw07} selection criteria.  The goal of this
paper is to compare the HST/WFC3 selected UV-dropout galaxies at
\uvdrops\ with similar galaxies at higher redshifts ($z\sim
4$--$5$). The limited area and depth of the ERS data puts brightness
limitations on our UV-dropout selection, which was confined to
comparatively brighter part of the rest-UV luminosity function (around the
knee and brighter as shown in H10). Therefore, we have
selected \acsb-dropouts and \acsv-dropouts in the ERS field whose
luminosities are similar to the UV-dropouts. The luminosity range is
$0.1L^*\lesssim L \lesssim 2.5L^*$ (based on $L^*$ corresponding to
M=--21 mag).  Applying similar filtering criteria as LBGs at \uvdrops,
we have a comparison sample of 155 \acsb-dropouts and 27
\acsv-dropouts in the ERS field.  Based on \citet{xue11} X-ray
catalog, there are four active galactic nuclei (AGN) in the H10 LBG
sample ($z\lesssim 3$), three AGN in the \acsb-dropout sample, and
none in the \acsv-dropout sample. These small number of X-ray AGN does
not affect our results or conclusions.

All subsequent analysis in this paper is done identically on these 5
samples (\wfcfuv-, \wfcnuv-, \wfcuv-, \acsb-, and \acsv-dropouts) for
proper comparison. To show general evolutionary trends, we combine
three dropout samples from H10 as a UV-dropout sample (\uvdrops), and
two high redshift samples as a \acsb-,\acsv-dropout sample (\bvdrops).


\section{Spectral Energy Distributions}\label{seds}

The \texttt{Le PHARE} software package \citep{arno99, ilbe06} was used
to measure the photometric redshifts, and to fit the broadband SEDs of
LBGs. The primary goal of SED fitting is to find the best-fitting
synthetic stellar population model to the observed photometry. From
this best-fit model, we can estimate the redshift, stellar age,
stellar mass, star-formation rate (SFR), dust extinction, and other
physical properties of each galaxy. We use the 2007 version of the
\citet[hereafter CB07]{bruz03} models, which has improved prescription
of thermally pulsating AGB stars. We generated a set of stellar
population models assuming a Salpeter initial mass function, and
varying the redshift ($z=0.1$--6.0, $\delta$z=0.1, though a parabolic
interpolation is used to refine the photometric redshift solution
within $\delta$z intervals), metallicity (0.2,
0.4 and 1 $Z_{\odot}$), age (1 Myr $\leq t \leq$ $t_{H}$), dust
extinction (0 $\leq$ E(B-V) $\leq$ 0.7~mag, using a modified
\citealt{calz00} attenuation law), and $e$-folding timescale
($\tau$=0.1,0.3,1,2,3,5,10,15,30~Gyr) for a star-formation history
(SFH)$\,\propto\,$exp(-t/$\tau$).  The \texttt{Le PHARE} code assumes
the \citet{mada95} prescription to estimate inter-galactic medium
(IGM) opacity. The model that gives the lowest $\chi^2$ is chosen as
the best-fit SED.

The contribution of major emission lines in different filters can be
included in the models using the \texttt{Le PHARE} code. Neglecting
emission lines during the SED fitting process can overestimate the
best-fit stellar ages and masses by as much as 0.3 dex
\citep[e.g.,][]{scha09,fink11,atek11}. The \texttt{Le PHARE} code
accounts for the contribution of emission lines with a simple recipe
based on the \citet{kenn98} relations between the SFR and UV
luminosity, H$\alpha$ and [OII] lines. The code includes the
Ly$\alpha$, H$\alpha$, H$\beta$, [OII], OIII[4959] and OIII[5007]
lines with different line ratios with respect to [OII] line, as
described in \citet{ilbe09}.

\begin{figure}
\begin{center}
\includegraphics[scale=0.68,angle=0]{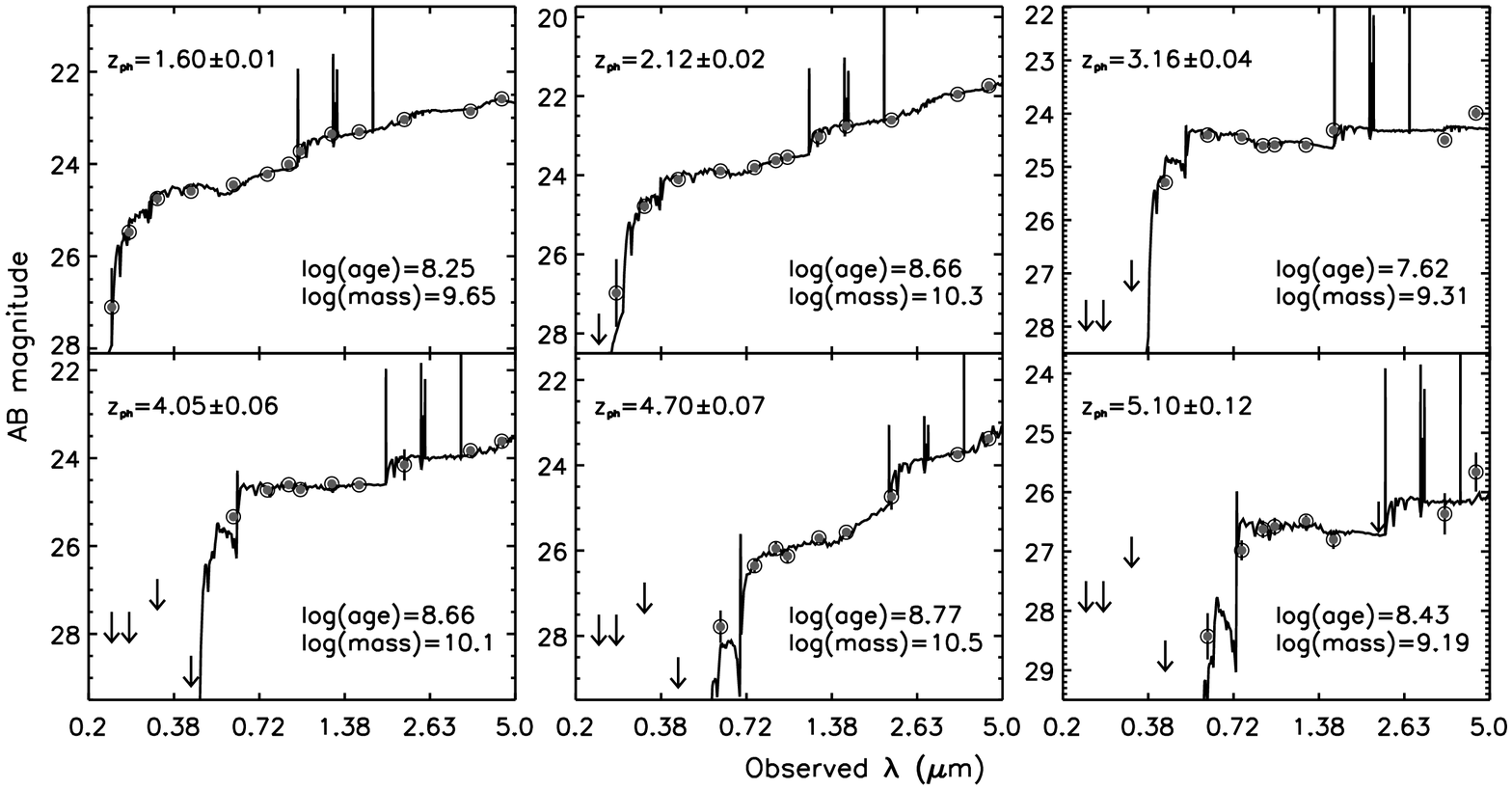}
\caption{[Top panel] Example best-fit SEDs of LBGs at \uvdrops. We
  have used the \texttt{Le PHARE} code \citep{arno99,ilbe06} to
  compute photometric redshifts and to perform SED fitting.  Grey
  concentric circles are observed magnitudes in three \emph{HST}/WFC3
  UVIS, four \emph{HST}/ACS, three \emph{HST}/WFC3 IR, one VLT Ks, and
  two \emph{Spitzer}/IRAC [3.6], [4.5] bands.  [Bottom panel] Same as
  the top panel but for the comparison sample of LBGs at
  \bvdrops. Stellar masses are in solar units
and stellar ages are in years.}\label{fig:seds}
\end{center}
\end{figure}

The observed photometry is available in up to 13 filters: three
\emph{HST}/WFC3 UVIS, four \emph{HST}/ACS, three \emph{HST}/WFC3 IR,
one VLT Ks, and two \emph{Spitzer}/IRAC [3.6], [4.5] bands. We perform
matched aperture photometry in 10 \emph{HST} bands as discussed in
H10, while we use VLT and \emph{Spitzer} photometry from the publicly
available GOODS-MUSIC catalog \citep{sant09}. The photometry in MUSIC
catalog has accurate PSF-matching of space and ground-based images of
different resolution and depth. \figref{fig:seds} shows example
best-fit SEDs for LBGs at \uvdrops, and the comparison sample of LBGs
at \bvdrops.

\subsection{Photometric Redshifts}\label{photz}

\begin{figure}
\begin{center} 
\includegraphics[scale=0.95,angle=0]{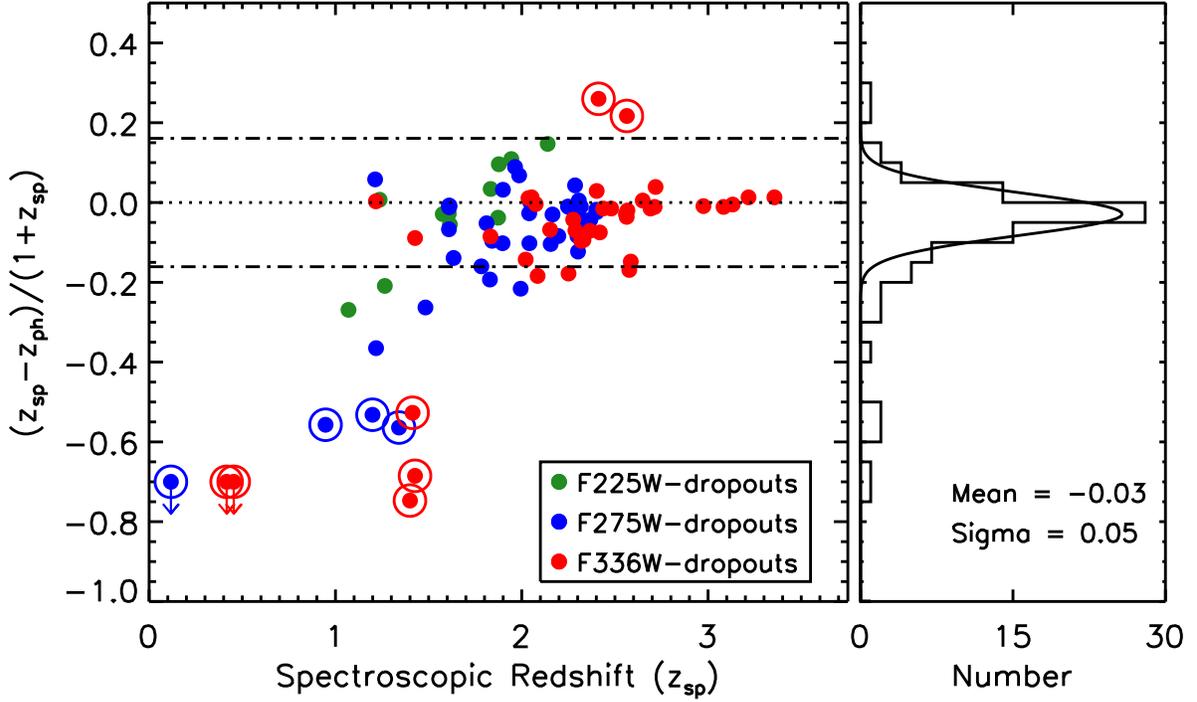}
\caption{Comparison of the SED-based photometric redshifts with the
  publicly available spectroscopic redshifts for LBGs at \uvdrops. The
  concentric circles indicate unreliable spectroscopic redshifts based
  on their quality flags. The dotted line shows the 1-to-1 relation
  between two redshifts. The histogram shows distribution of ($z_{sp}
  - z_{ph}$/1+$z_{sp}$), and the 1$\sigma$ uncertainties in this
  distribution is $\sim$5\%.  The dot-dash lines show 3$\sigma$
  limits.}\label{fig:redshifts}
\end{center}
\end{figure}

One of the free parameters during the SED fitting process is the
redshift.  To assess the accuracy of our SED-based photometric
redshifts ($z_{ph}$) at \uvdrops, we compare them with the
spectroscopic redshifts ($z_{sp}$) from various VLT/Magellan campaigns
in the GOODS-S field
\citep[e.g.,][]{graz06,ravi07,vanz08,wuyt08,bale10,coop12}.  We find
that only a small number ($\lesssim\,$30\%) of H10 dropout sample has
spectroscopic redshifts, most likely due to the lack of strong
features in observed 4500--9000~\AA\ range at $1 \lesssim z \lesssim
3$, where most ground-based spectrographs on large telescopes are
optimized.  We matched 91 spectroscopic redshifts for the UV-dropout
sample (\uvdrops) selected with the criteria discussed in
\secref{data}. \figref{fig:redshifts} shows the comparison between the
SED based photometric redshifts and the publicly available
spectroscopic redshifts.  The catastrophic outliers --- shown by
concentric circles in \figref{fig:redshifts} --- have quality flags
that indicate the spectroscopic redshift is unreliable (in most
catalogs C or worst). So it is likely that these spectroscopic
redshifts are not correct and hence, redshift comparison for these
objects is not credible.  The histogram in \figref{fig:redshifts}
shows the distribution of photometric redshift uncertainties $\delta
z$=($z_{sp} - z_{ph}$/1+$z_{sp}$).  Based on this distribution, we
estimate $\sigma$($\delta z$)$\simeq$0.05, and $<\delta
z>$=--0.03. The fraction of catastrophic outliers ($>$3$\sigma$) is
$\sim$7\%, excluding objects with unreliable spectroscopic redshifts.
Our photometric redshift uncertainties are consistent with
\citet{dahl10}, who used the deepest and the most comprehensive
photometric data in the GOODS-S field. \citet{habe12} selected fairly
bright LBGs at $z\simeq 2$ using the \emph{GALEX} data, and found
similar photometric redshift uncertainties and outlier fraction for
their dropout sample. The photometric redshift uncertainty in the
implied redshift is also consistent with the dropout selection method
applied to select these galaxies. The dropout selection technique uses
the location of a spectral break within a photometric bandpass
(filter), and therefore, the redshift uncertainty depends on the width
of the bandpass, and could be as high as $\sim$0.5 in $z$.

The distribution of photometric redshift uncertainties ($\delta z$) is
asymmetric, even after excluding objects with unreliable spectroscopic
redshifts. There are more galaxies in the distribution with 
spectroscopic redshifts lower than their photometric redshifts i.e.,
($z_{sp} - z_{ph} < 0$). Detail investigation of each ground-based
spectra (if available) is needed to figure out what is causing
spectroscopic redshift to be lower than photometric redshift.  Such an
investigation is beyond the scope of this paper, but we should point
out that such asymmetric distribution is also observed for
\emph{GALEX}--selected LBGs \citep[e.g.,][]{habe12}, and is totally
consistent within the estimated photometric redshift uncertainties.

\subsection{UV Spectral Slope $\beta$}\label{beta}

\begin{figure}
\begin{center}  
\includegraphics[scale=0.88,angle=0]{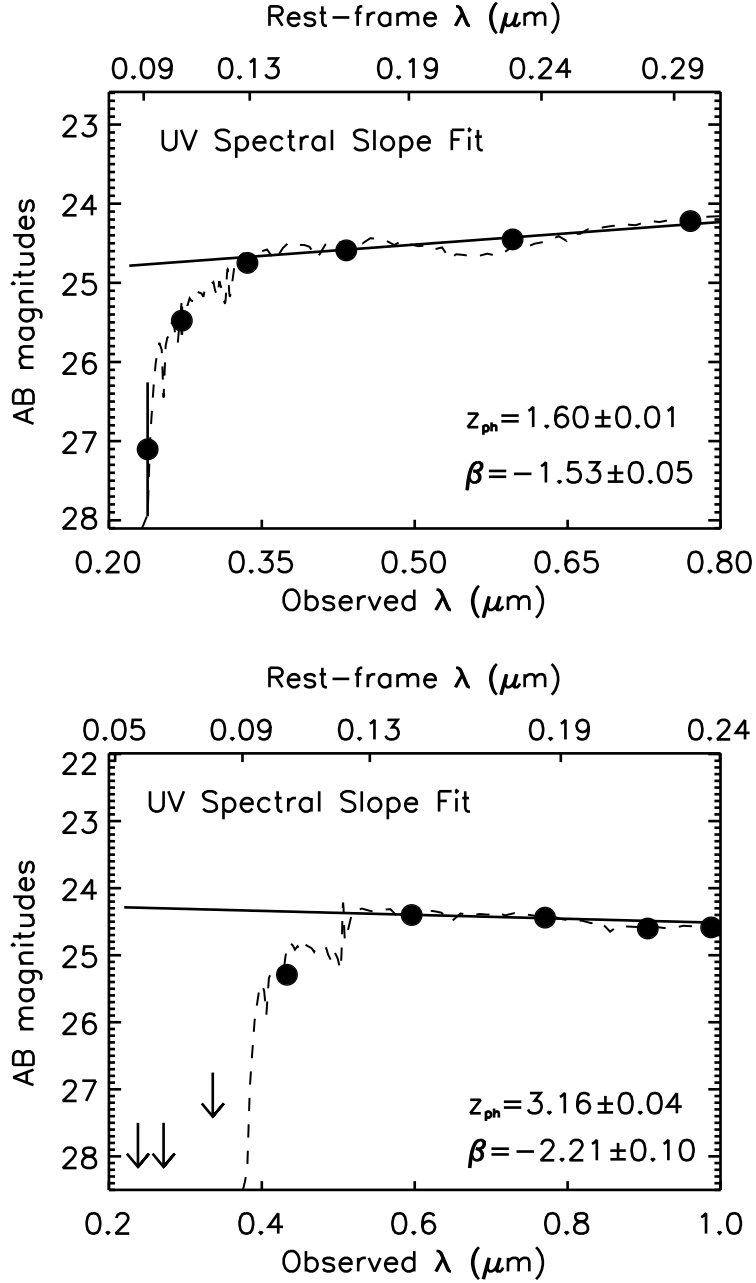}
\caption{Example fitting of the UV spectral slope $\beta$ to best-fit
  SEDs for two LBGs from the \uvdrops\ sample. The dotted black
  line is the best-fit SED, while solid black line shows the
  estimated UV continuum slope $\beta$. The black filled circles are observed
  magnitudes. We test two different rest-frame UV wavelength range
  (1300--1900~\AA\ and 1300--3400~\AA) to address the robustness
  of our estimated UV continuum slope $\beta$, and find similar median
slopes within 1$\sigma$ scatter.}\label{fig:beta_fit}
\end{center}
\end{figure}

The UV spectral slope $\beta$ is determined from a power-law fit to
the UV continuum spectrum \citep{calz94}, $f_{\lambda} \varpropto
\lambda^{\beta}$, where $f_{\lambda}$ is the flux density per unit
wavelength (ergs s$^{-1}$ cm$^{-2}$ \AA$^{-1}$). We use the best-fit
SEDs of dropout selected LBGs to estimate their UV spectral slope
$\beta$ by fitting a straight line between rest-frame 1300 and
1900~\AA\ in their model spectrum. This wavelength range covers 7 out
of 10 spectral fitting windows identified by \citet{calz94} to
estimate the UV spectral slope. This wavelength range is also ideal
for comparing $\beta$ values at higher redshifts, because those are
usually measured between rest-frame 1600 and
2000~\AA. \figref{fig:beta_fit} shows the slope-fitting method applied
to the best-fit SEDs to estimate $\beta$, where the solid line is the
best-fit UV spectral slope, the dashed line is the best-fit SED, and
the black filled circles are observed magnitudes.  By selection
(\secref{data}), we only consider galaxies with good SED fits so the
choice of model should not affect the $\beta$ estimate.  Though
uncertainties in observed photometry could affect the best-fit SED
parameters, and hence, the $\beta$ estimate. In \figref{fig:beta_fit}
, we also quote typical intrinsic uncertainty in $\beta$ for galaxies
at different redshifts.  We estimate $\beta$ for each galaxy, and then
fit a Gaussian to the $\beta$ distribution to find median (and sigma)
value in each redshift bin. \tabref{tab:beta} shows median $\beta$
values and their corresponding uncertainties for the UV-dropout, and
the \acsb-,\acsv-dropout samples.

\begin{figure}
\begin{center}  
\includegraphics[scale=0.88,angle=0]{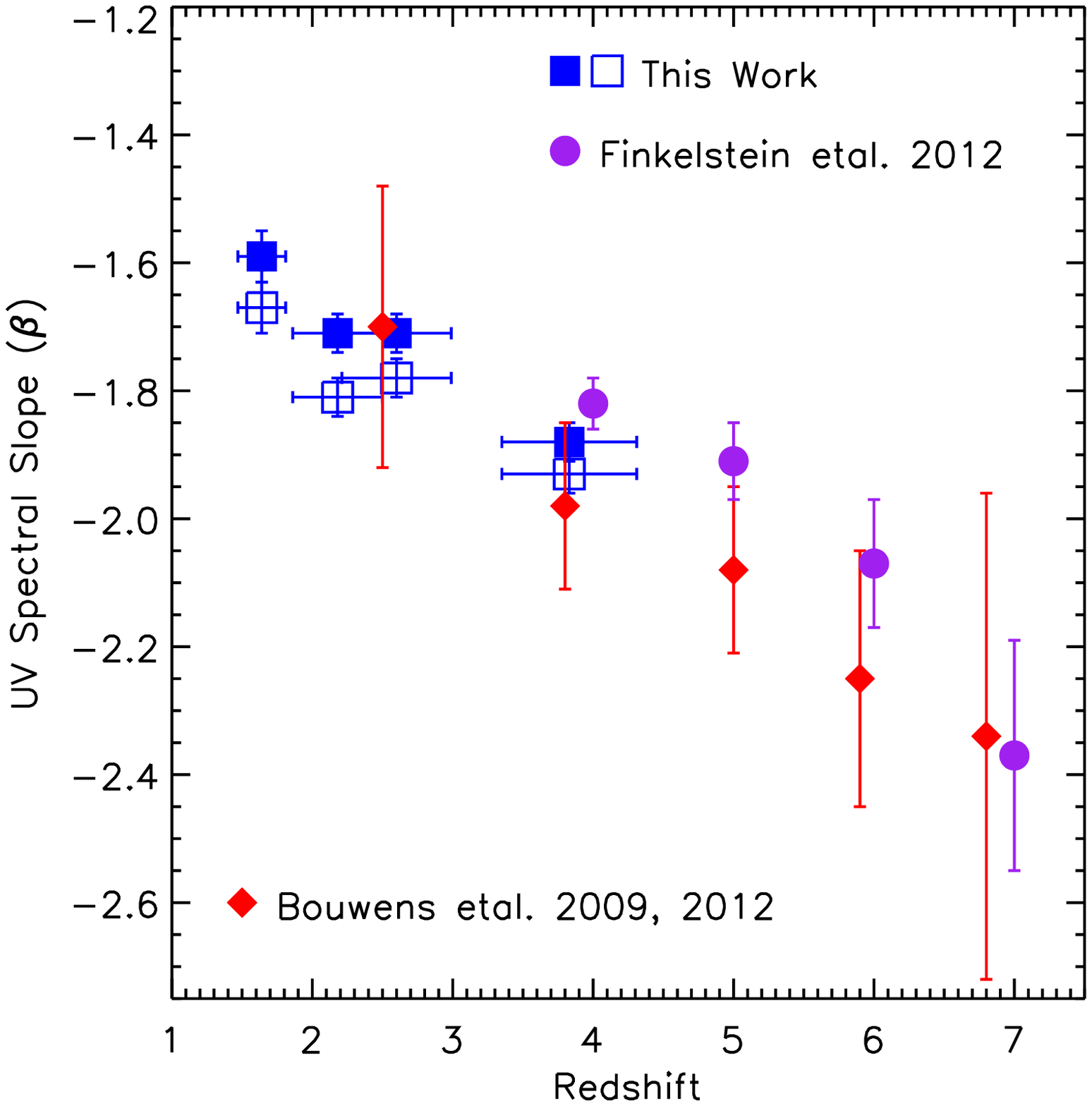}
\caption{UV spectral slope $\beta$ as a function of redshift. Blue
  filled squares indicate the median $\beta$ values when we fit
  rest-frame 1300 to 1900~\AA\ wavelengths, and blue open squares
  indicate $\beta$s when we fit 1300 to 3400~\AA. The
  higher redshift ($z\gtrsim 2.5$) measurements from \citet{fink12}
  and \citet{bouw09,bouw12} are also shown for comparison.  Ours and the
  \citet{fink12} uncertainties are the standard error of the mean,
  while uncertainties from Bouwens et al. are 1$\sigma$ scatter. For
  comparison, our 1$\sigma$ uncertainties/scatter are listed in
  \tabref{tab:beta}.}\label{fig:beta_z}
\end{center}
\end{figure}

The evolution in the UV spectral slope $\beta$ as a function of
redshift may indicate change in stellar populations of galaxies over
cosmic time. We compare our $\beta$ values with the higher redshift
measurements from the literature \citep[e.g.,][]{bouw12,
  fink12}. \figref{fig:beta_z} shows the UV spectral slope $\beta$ as
a function of redshift. Blue filled squares are median $\beta$ values
measured between rest-frame 1300 and 1900~\AA\ for our dropout
samples.  To test how $\beta$ measurements are affected by the
selection of rest-frame UV wavelength range, we also measured $\beta$
between rest-frame 1300 and 3400~\AA, which are shown by blue open
squares in \figref{fig:beta_z}. Both $\beta$ values agree within
1$\sigma$ uncertainties.  The red filled diamonds are measurements
from \citet{bouw09,bouw12} and purple filled circles are from
\citet{fink12}. The $\beta$ values from \citet{bouw09,bouw12} are for
the galaxies around $M_{\rm uv}^*$, which is consistent with our
sample, while \citet{fink12} measurements are based on all galaxies
extending to those fainter than $M_{\rm uv}^*$ in their respective
redshift bins. The uncertainties on the median values of $\beta$ are
the standard error of the mean in the case of \citet{fink12} and our
measurements, while \citet{bouw09,bouw12} uncertainties represent
1$\sigma$ scatter.  For comparison, our estimated 1$\sigma$ scatter in
median $\beta$ values are listed in \tabref{tab:beta}.

\begin{figure}
\begin{center}  
\includegraphics[scale=0.44,angle=0]{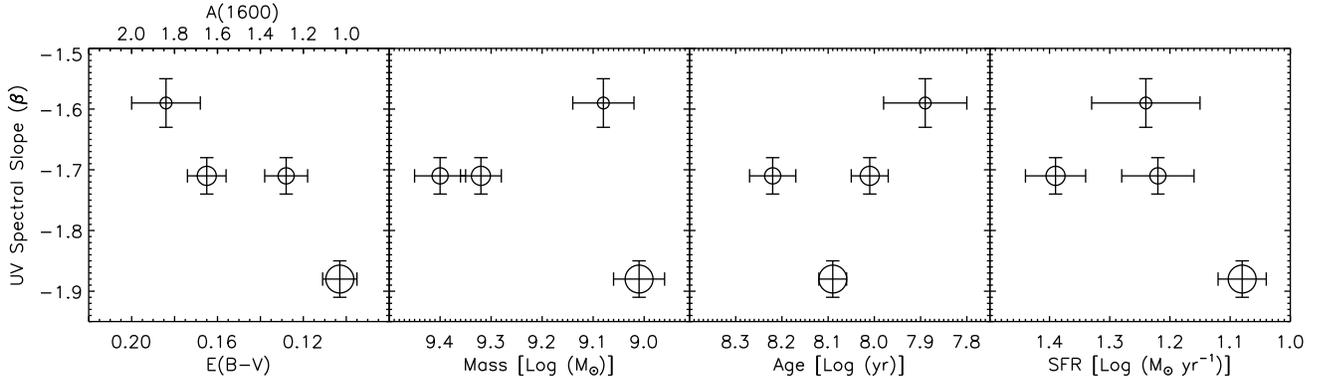}
\caption{UV spectral slope $\beta$ versus best-fit SED parameters
  (E(B-V), stellar mass, stellar age, star-formation rate). The
  size of the circle increases with redshift i.e.,  the smallest
  circle correspond to the lowest redshift bin ($z\simeq 1.6$), 
  and the largest circle correspond to  $z\simeq 4$--$5$ bin. The changes
  in the UV-continuum slope $\beta$ from $z\simeq 1.6$ to $5$ are most
  likely due to the change  (factor of $\sim$2 or 0.3 dex) in the dust
  content E(B--V) of the galaxies, as other parameters show smaller
  variation as a function of
  redshift.}\label{fig:beta_all}
\end{center}
\end{figure}

\figref{fig:beta_z} shows that the median values of $\beta$ decreases
as redshift increases ($\beta\simeq\,$--1.6 at $z\simeq 1.6$ to
$\beta\simeq\,$--2.4 at $z\simeq 8$), which could imply variations in
one or more physical properties of LBGs as a function of redshift.
\figref{fig:beta_all} shows evolution in $\beta$ as a function of
best-fit SED parameters (stellar mass, stellar age, dust content, SFR)
and redshift. The lowest redshift bin ($z\simeq 1.6$) is shown by the
smallest circles, and the highest redshift bin ($z\simeq 4$--$5$) is
shown by the largest circles. The largest change (factor of $\sim$2 or
0.3 dex) is seen in the dust content E(B--V) of galaxies as
$\beta$ changes from --1.6 (at $z\simeq 1.6$) to --1.9 (at $z\simeq
4$--$5$), while other three parameters vary much less than a factor of
2. This could imply that change in the dust content of LBGs has
largest effect on the UV spectral slope $\beta$, and any variation in
$\beta$ as a function of redshift could most likely be due to changing
dust content of galaxies. Therefore, based on our $\beta$ estimates,
as shown in \figref{fig:beta_z}, we could say that LBGs at lower
redshift ($z\simeq 1.6$) have more dust than LBGs at higher redshift
($z\simeq 5$).  This trend of $\beta$ is consistent with 
previous studies, which argued that galaxies at $z\simeq 6$ tend
to be bluer than those at $z\simeq 3$
\citep[e.g.,][]{stan05,bouw06,hath08a,wilk11}.  Those $\beta$
measurements, on uniformly selected LBGs, were limited to LBGs at
$z\gtrsim 3$, and our results below $z\simeq 3$ extends this observed
trend to $z\simeq 1.5$.

The evolution in the UV spectral slope $\beta$ could also be due to
changing star-formation history, initial mass function (IMF), and/or
metallicity.  These effects are believed to be much smaller than the
effects from changing dust content of the galaxy.  Many authors have
investigated various stellar population models to estimate these
effects. \citet{bouw12} explored sensitivity of the UV-continuum slope
$\beta$ to changes in the mean metallicity, age, or dust extinction by
choosing one fiducial model as a benchmark, and then changing various
model parameters to assess changes in $\beta$. They conclude that a
factor of 2 (or 0.3 dex) changes in metallicity, age or dust content
result in 0.07, 0.15, 0.35 changes in the UV spectral slope $\beta$,
respectively. This implies that changes in the dust content have much
larger effect on the UV-continuum slope than similarly-sized changes
in the age, metallicity, or the stellar IMF. Similar studies
\citep[e.g.,][]{leit99,hath08a, wilk11} have come to the same
conclusion that though $\beta$ could be affected by various stellar
population properties, the change in dust content of galaxies is the
predominant effect which causes $\beta$ to change. We should also
emphasize that it is very challenging to completely understand these
various effects based on observations only, rigorous modeling and/or
simulations are required to fully assess the contributions of these
various effects on the UV spectral slope $\beta$.

\subsection{Stellar Population Properties}\label{age_mass_sfr}

We compare observed SED with a suite of model templates from CB07 to
find the best-fit model through $\chi^2$ minimization.  All SED 
parameters are fit simultaneously. The best-fit
model allows us to estimate photometric redshift (as shown in
\secref{photz}), and physical properties of stellar populations such
as stellar age, stellar mass, dust extinction E(B--V) and SFRs for
each galaxy. The estimated uncertainties ($\sim$0.3--0.4 dex) in
stellar ages, masses and SFRs are estimated by marginalizing the
uncertainties in observed photometry and redshift.

One of the main limitations of SED fitting is the need to assume a
SFH, which cannot be reliably constrained for a galaxy from limited
photometric data points. We have assumed an exponentially declining
SFH. Different SFHs (e.g., rising, constant, declining) introduce
systematic uncertainties in the stellar mass determinations, mostly at
redshift greater than $z\simeq 3$ \citep[e.g.,][]{lee10,papo11}.
These uncertainties are typically $\lesssim$0.3 dex
\citep[e.g.,][]{finl07}, and are within our estimated uncertainties.
Stellar ages are highly sensitive to the assumed SFH. Any prior
star-forming episode can be overshadowed by newly born stars from the
most recent star-formation, totally neglecting possible existence of
older population in a given galaxy. Therefore, based on assumed
histories, it is possible to get an older or a younger age for the
same galaxy. Hence, interpreting the stellar ages derived from SED
fitting can be tricky, and at the very least, the uncertainties on the
stellar ages could be much larger than estimated uncertainties
($\sim$0.3--0.4 dex).  Because of these issues, in subsequent
analysis, we will not elaborate on stellar population ages and focus
on other physical properties.

\begin{figure}
\begin{center}  
\includegraphics[scale=0.8,angle=0]{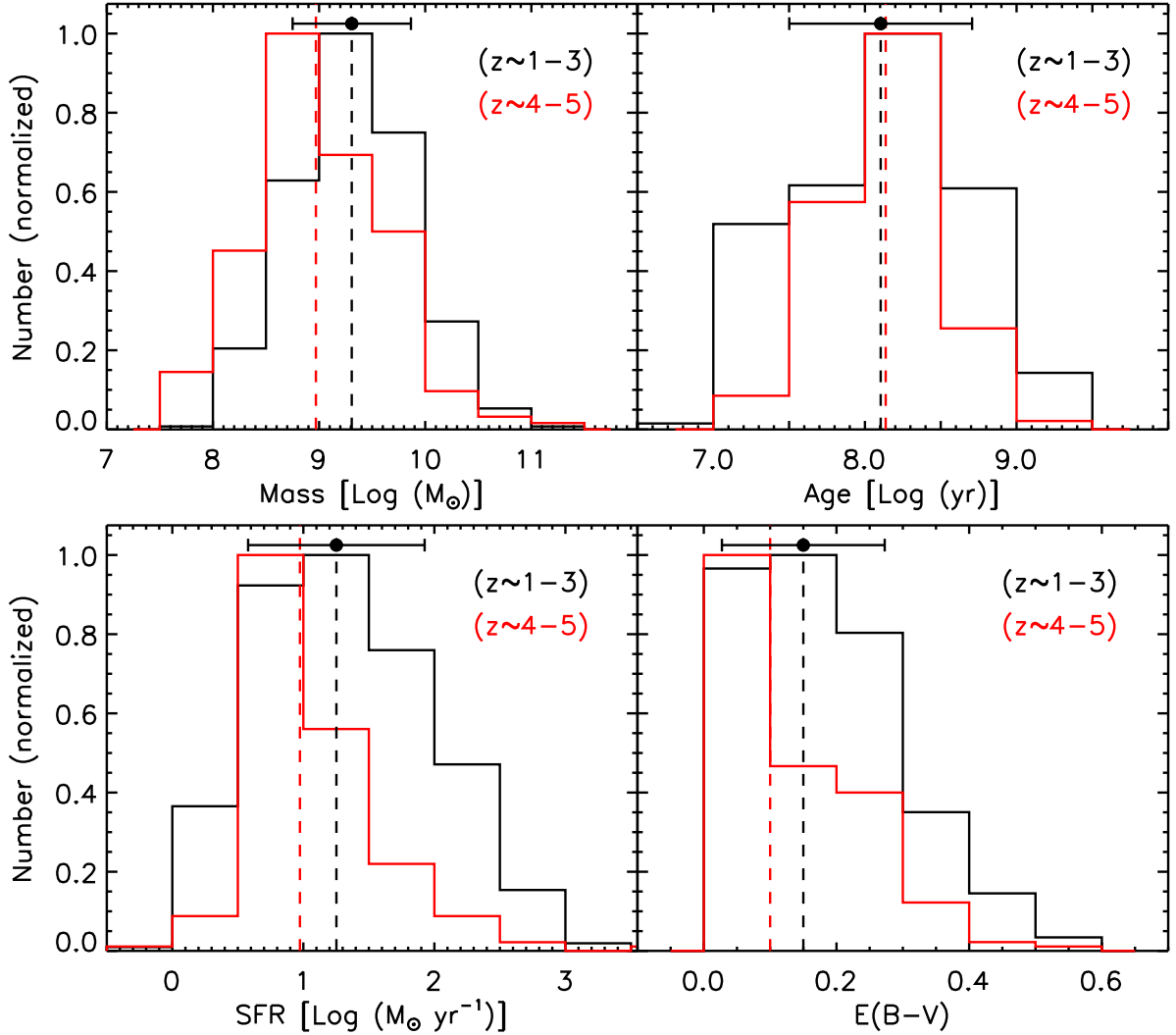}
\caption{Distribution of stellar masses, stellar ages, star-formation
  rates, and dust extinction for LBGs at \uvdrops\ (black) and the
  comparison sample at \bvdrops\ (red). All four parameters are
  estimated from the best-fit SEDs. Median values are shown by
  vertical dashed lines and 1$\sigma$ uncertainties are shown by an
  error bar at the top of the black histogram. Though the higher
  redshift LBGs have, on average, lower values of most of these
  physical parameters than LBGs at \uvdrops, the median values are
  similar within estimated uncertainties. Stellar ages do not show any
  clear evolution with redshift mainly because of higher uncertainties
  in their measurements (see \secref{age_mass_sfr}). The histograms
  show normalized numbers for both samples (total number of galaxies
  in each sample is shown in \tabref{tab:beta}), and a two-sided K-S
  test --- in each panel --- indicates a probability $P\lesssim 0.006$
  that the two distributions are drawn from the same parent
  distribution.}\label{fig:sed_hist}
\end{center}
\end{figure}

\figref{fig:sed_hist} shows the distributions of stellar age, stellar
mass, SFRs, and E(B--V) for LBGs at \uvdrops\ (black), and the
comparison sample of LBGs at \bvdrops\ (red).  The median values are
shown by dashed vertical lines and 1$\sigma$ uncertainties in these
distributions for LBGs at \uvdrops\ are shown by an error bar at the
top of the black histogram.  A two-sided K-S test --- in each panel
--- indicates a probability \emph{less} than 0.006 that the
distributions (red and black histograms) are drawn from the
\emph{same} parent distribution.  \figref{fig:sed_hist} shows a
general trend that --- on average --- higher redshift LBGs have low
SFRs, less dust, and are less massive than their lower redshift
counterparts, though median values of two distributions (red and
black) are similar within 1$\sigma$ uncertainties.  This result is in
good agreement with previous studies comparing LBGs at $z\simeq 3$ and
$z\simeq 5$ \citep[e.g.,][]{verm07}. The average E(B--V) at $z\simeq
2$ is consistent with studies based on star-forming galaxies selected
using BX/BM color technique \citep[e.g.,][]{erb06, sawi12}.  The
distribution of E(B--V) completely agrees with the UV spectral slope
evolution as discussed in \secref{beta}, implying that the LBGs at
\uvdrops\ are more dusty (redder) compared to LBGs at \bvdrops.


\section{Results and Discussion}\label{results}

\subsection{Stellar Mass vs UV Luminosity Relation}\label{mass_lum}

\begin{figure}
\begin{center}  
\includegraphics[scale=0.8,angle=0]{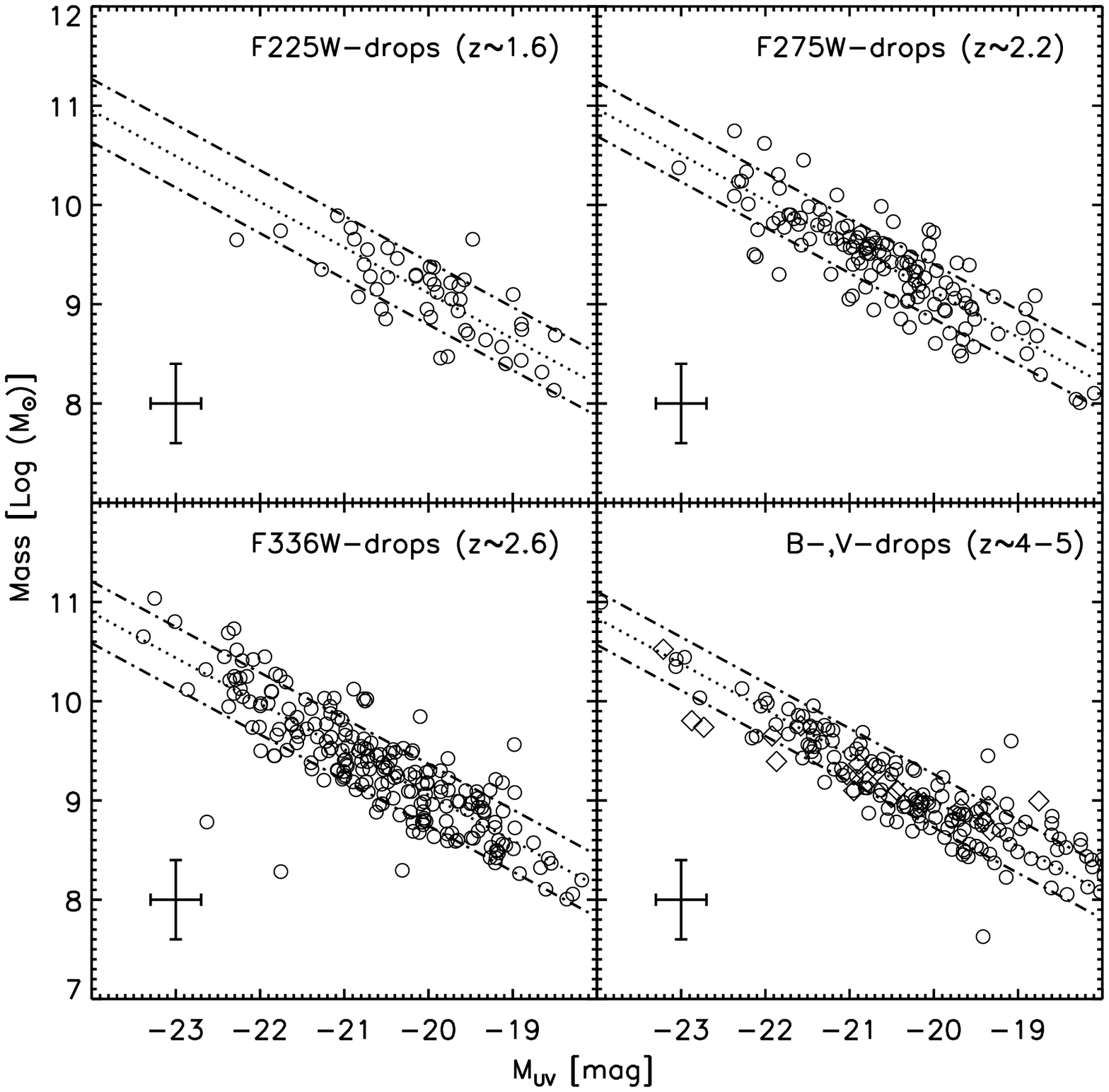}
\caption{Stellar mass versus UV absolute magnitude relation for
  \wfcfuv-, \wfcnuv-, \wfcuv-dropout samples. Bottom right panel shows
  the same relation for the comparison sample of LBGs at \bvdrops\
  (black circles for $z\simeq 4$, and black squares for $z\simeq
  5$). The average uncertainties are shown in the bottom-left
  corner. The dotted black line is the best-fit line with a
  logarithmic slope of 0.46. The dot-dash line shows the 1$\sigma$
  scatter ($\sim$0.3 dex) from the best-fit linear
  relation.}\label{fig:mass_abs}
\end{center}
\end{figure}

The rest-frame UV light traces recent or instantaneous SFR, while
rest-frame optical and NIR data help us to estimate stellar masses of
galaxies. If the galaxy stellar mass and UV luminosity are related
then we can directly use rest-frame UV light to estimate stellar mass
without needing rest-frame optical/NIR data.  \figref{fig:mass_abs}
shows stellar mass of LBGs at $z\simeq 1.5$--$5$ as a function of
their UV absolute magnitude.  These quantities are based on best-fit
SEDs, and their typical uncertainties are shown in the lower-left
corner. The dotted lines are best-fit line obtained by keeping the
logarithmic slope fixed at 0.46, which was estimated by \citet{sawi12}
for star-forming galaxies at $z\simeq 2$.  The dot-dash lines show the
scatter from the best-fit line, which is $\sim$0.3 dex for LBGs at
\uvdrops\ and about 0.2 dex for LBGs at \bvdrops. We also tested the
validity of this relation by fitting the slope of the line rather than
fixing it. We find that the fitted slope is in the range of
0.42$\pm$0.06 for our LBG samples, which is consistent with 0.46
within the estimated 1$\sigma$ scatter in this relation. Therefore, we
find that a proportionality relation between these two parameters with
a logarithmic slope of 0.46 provides a good fit to the data. The
stellar masses of the brighter LBGs --- with UV luminosities near the
$L_{\rm uv}^*$ value of LBGs at $z\simeq 3$ from \citet{stei99} ---
are about a factor of 2 lower than 10$^{10}$~M$_{\odot}$ estimated by
\citet{papo01}. This discrepancy, though within our estimated
uncertainties, could be due to the fact that we include emission lines
in our SED fitting which could affect stellar masses by as much as a
factor of $\sim$2. The stellar mass--UV luminosity relation is fairly
tight with a small scatter ($\lesssim\,$0.3 dex), which is consistent
with other studies at similar redshifts \citep[e.g.,][]{papo01}, and
it points to a nearly constant mass-to-light ratio
(log(M/L)$\,\simeq\,$--0.5) for LBGs between $z\simeq 1.5$ and $5$.
The tightness/lower scatter of the stellar mass--UV luminosity
relation in \figref{fig:mass_abs} could be --- in part --- due to the
fact that both these quantities are output parameters from the
best-fit SEDs, and therefore, it is possible that these parameters are
not totally independent. We have addressed this issue and discussed
its implications in \secref{discuss}.  A similar correlation between
stellar mass and absolute magnitude has been reported for LBGs at
$z\simeq 5$--$6$ by \citet{star09}.

\figref{fig:mass_abs} shows that LBGs at $z\simeq 1.5$--$5$ follow
similar linear correlation between stellar mass and UV absolute
magnitude (within uncertainties) for $M_{\rm uv}$ between --19 and
--22.5~mag.  It is important to note that \citet{shap05} does not find
any correlation between the stellar mass and UV absolute magnitude for
star-forming galaxies at $z\simeq 2$ with stellar masses
$\gtrsim\,$10$^{10}$~M$_{\odot}$. This could be due to different
color-selection technique (BX/BM) used by the \citet{shap05} to select
star-forming galaxies, whose physical properties could differ from the
dropout selected LBGs at these masses \citep[e.g.,][]{ly11,habe12}. It
is also possible that their sample --- which consists of
spectroscopically confirmed bright galaxies with stellar masses
greater than or equal to 10$^{10}$~M$_{\odot}$ --- has more massive
galaxies than our sample and it is uncertain how massive galaxies
would follow this correlation.  The ERS observations are too limited
in area and depth to cover a larger luminosity range, so we cannot
predict how this relation will evolve for luminous ($M_{\rm uv}\!
<\,$--22.5~mag) or dwarf ($M_{\rm uv}\!>\,$--19~mag) galaxies at these
redshifts.

\subsection{SFR vs Stellar Mass}\label{sfr_mass}

\begin{figure}
\begin{center}  
\includegraphics[scale=0.8,angle=0]{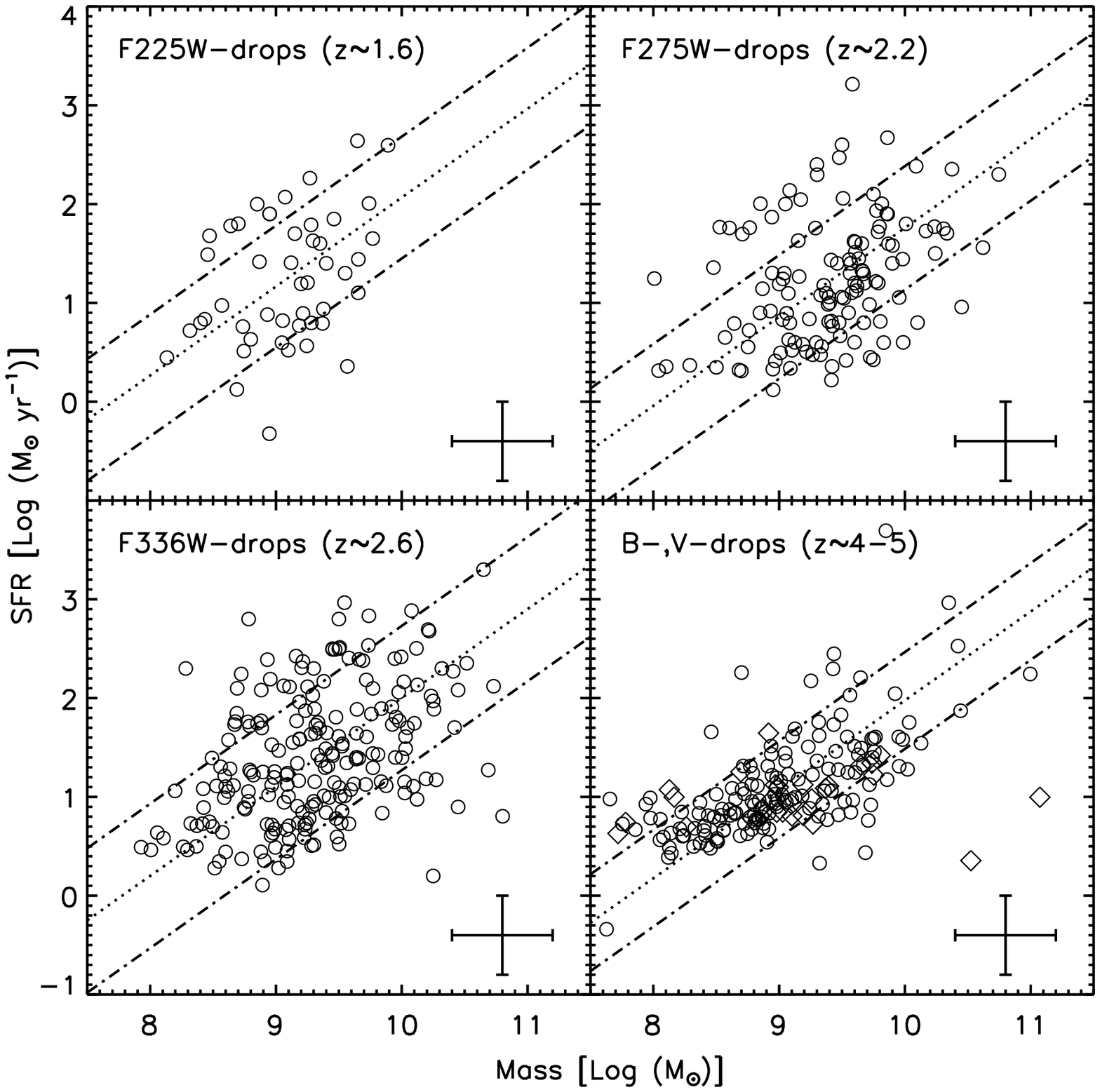}
\caption{ Stellar mass versus SFR relation for \wfcfuv-, \wfcnuv-,
  \wfcuv-dropout samples. Bottom right panel shows the same relation
  for the comparison sample of LBGs at \bvdrops\ (black circles for
  $z\simeq 4$, and black squares for $z\simeq 5$). The average
  uncertainties are shown in the bottom-right corner. The dotted black
  line is the best-fit line with a logarithmic slope of 0.90. The
  dot-dash line shows the 1$\sigma$ scatter ($\sim$0.6 dex, $\sim$0.4
  dex for \bvdrops) from the best-fit linear relation. In the upper
  panel, few galaxies form a sharp edge towards high SFR values, which
  could be an artifact due to lower limits on the model parameters
  $\tau$ and \emph{t} \citep[e.g.,][]{hain12}.  }\label{fig:mass_sfr}
\end{center}
\end{figure}

The correlation between the current SFR and stellar mass in
star-forming galaxies, also known as `main sequence of star-formation'
(MS), has been observed at $z\lesssim 2$
\citep[e.g.,][]{noes07,elba07,dadd07}. These studies have shown that
the MS relation seems to be not evolving strongly with redshift, but
the zeropoint does: that is high redshift ($z\simeq 2$) star-forming
galaxies are forming stars at a higher rate than similar mass local
galaxies. In \figref{fig:mass_sfr}, we investigate this relation and
star-formation histories for LBGs at $z\simeq 1.5$--$5$.  These
quantities are based on best-fit SEDs, and their typical uncertainties
are shown in the lower-right corner. The dotted lines are best-fit
line obtained by keeping the logarithmic slope fixed at 0.90,
estimated for star-forming galaxies at $z\le 2$
\citep[e.g.,][]{elba07,dadd07,sawi12}. The dot-dash lines show the
scatter from the best-fit line, which is $\sim$0.6 dex for LBGs at
\uvdrops\ and about 0.4 dex for LBGs at \bvdrops. We also obtained the
best-fit logarithmic slope for this relation, and found the fitted
slope in the range of 0.81$\pm$0.30 for our LBG samples, which is
consistent with 0.90 within the estimated 1$\sigma$ scatter in this
relation.  We find that a proportionality with a logarithmic slope of
0.90 provides a good fit to the data with few outliers at stellar mass
greater than 10$^{10}$~M$_{\odot}$.

\citet{finl06} have shown that tight relation exists between SFR and
stellar mass for galaxies at $z\simeq 4$ using the cosmological
hydrodynamic simulations, which is also consistent with the
observations \citep{bouw12}.  \citet{finl06} also point out that the
scatter in the \figref{fig:mass_sfr} could be a measure of SFR
`burstiness' as a function of stellar mass. This means that the linear
relation (with a logarithmic slope of $\sim$0.90) indicate an average
SFR for a given stellar mass, but galaxies can also experience bursts
of up to two times the average SFR value at the same stellar mass as
shown by the scatter.  The scatter in the SFR versus stellar mass
relation for LBGs at \uvdrops\ is slightly larger than $\sim$0.3 dex
--- observed at $z\simeq 2$ by \citet{dadd07} --- possibly because of
few galaxies forming a sharp edge towards high SFR values, as seen in
the relation for \wfcfuv- and \wfcnuv-dropouts (upper panel in
\figref{fig:mass_sfr}).  These galaxies have low stellar ages (less
than 10~Myr), which could be highly uncertain as discussed in
\secref{age_mass_sfr}. It is also possible that this edge
could be an artifact due to lower limits on the model parameters $\tau$
and \emph{t} \citep[e.g.,][]{hain12}.  We also note that
\citet{mclu11} argue that the tightness in the SFR-stellar mass
relation depends on the assumed SFH. The scatter in this relation is
much less for a constant SFH, while it is much larger for other
SFHs. Therefore, it is also likely that the larger scatter we see in
\figref{fig:mass_sfr} could be due to different SFHs.

\figref{fig:mass_sfr} shows that, though our data has little more
scatter compared to the MS relation at $z\lesssim 2$, the majority of
our galaxies fall on to this relation characterized by a logarithmic
slope of 0.90. A similar correlation is observed at $z\simeq 6$--$8$
by \citet{mclu11}, and supported by cosmological hydrodynamic
simulations of \citet{finl11}. Our observations confirm this MS
relation for star-forming galaxies from $z\simeq 1.5$ to $5$, implying
that --- on average --- their star-formation histories are similar.

\subsection{Implications}\label{discuss}

In previous sections, we have shown that LBGs at \uvdrops\ --- on
average --- are massive, dustier, and have higher star-formation rates
than LBGs at \bvdrops\ with similar luminosities, though it should
also be noted that they are not very different within estimated
1$\sigma$ uncertainties. As pointed out by \citet{papo11}, the number
densities of galaxies at fixed luminosity could change substantially
over this redshift range, which could lead to potential biases when
comparing galaxies at different redshifts.  However, the general
trends we observe in stellar masses, SFRs, and dust extinction are
supported by other independent means. The characteristics UV
luminosity ($L_{\rm uv}^*$) is increasing as a function of redshift
from $z\sim 8$ to $2$ (e.g., H10), which implies increase in SFRs with
time, while \citet{fink10} have shown that stellar masses for $M_{\rm
  uv}^*$ LBGs grow from $z\simeq 8$ to $2$. The UV spectral slope
$\beta$ shows evolution as a function of redshift
(\figref{fig:beta_z}), which could indicate lower dust content at
higher redshifts.  The higher dust content in LBGs at lower redshift
is also in accordance with the studies at $z\simeq 1$
\citep[e.g.,][]{burg07,basu11}, while the \citet{verm07} supports the
lower dust content in LBGs at $z\simeq 5$. Therefore, the ensemble
properties of LBGs in our sample are in general agreement with the
expected results.

\begin{figure}
\begin{center}  
\includegraphics[scale=0.75,angle=0]{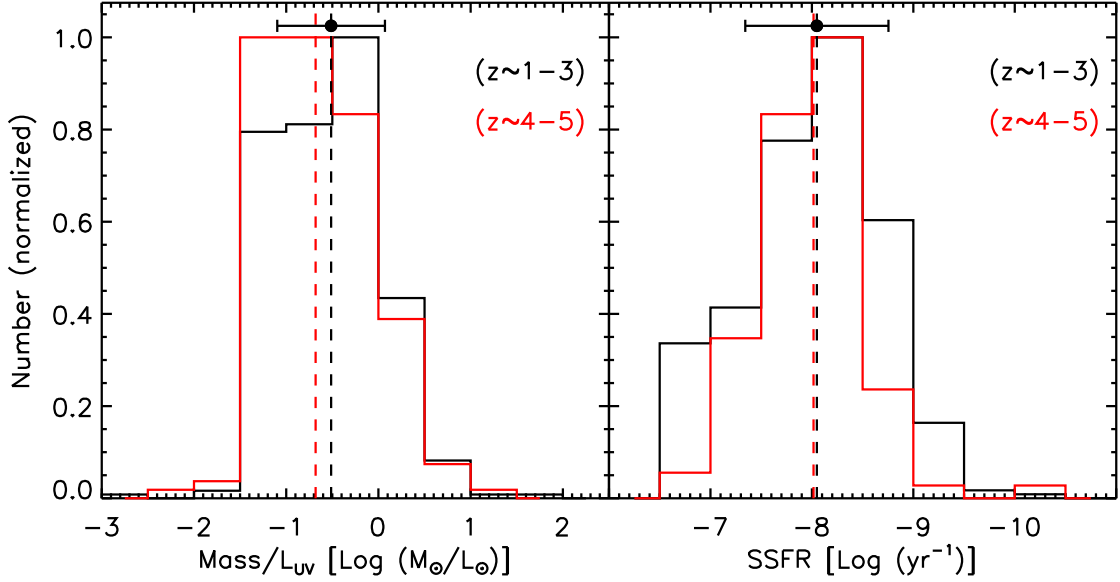}
\caption{Distribution of mass-to-light ratios and specific SFRs for
  LBGs at \uvdrops\ (black) and for the comparison sample at \bvdrops\
  (red). Median values are shown by vertical dashed lines and
  1$\sigma$ uncertainties are shown by an error bar at the top of the
  black histogram. Though the higher redshift LBGs have, on average,
  lower values of these parameters than LBGs at \uvdrops, the median
  values are similar within estimated uncertainties. The histograms
  show normalized numbers for both samples (total number of galaxies
  in each sample is shown in \tabref{tab:beta}), and a two-sided K-S
  test --- in each panel --- indicates a probability $P\lesssim 0.05$
  that the two distributions are drawn from the same parent
  distribution.}\label{fig:ssfr}
\end{center}
\end{figure}

The stellar mass--UV luminosity relation (\figref{fig:mass_abs}) and
the SFR--stellar mass relation (\figref{fig:mass_sfr}) are based on
measurements from best-fit SEDs, therefore, it is possible that these
quantities are not totally independent, which might affect their
observed correlations. To investigate this, we show distributions of
mass-to-light (M/L; Mass/$L_{\rm uv}$) ratios and specific SFRs (SSFR;
SFR/Mass) in \figref{fig:ssfr}. The black (red) histograms show
distribution for LBGs at \uvdrops\ (\bvdrops), and the median values
are shown by dashed vertical lines. The median values of M/L ratio and
SSFR for LBGs at \bvdrops\ are slightly lower than that at \uvdrops,
but are still consistent within the 1$\sigma$ uncertainties as shown
by the error bar on the top of the black histogram.  A two-sided K-S
test --- in each panel --- indicates a probability \emph{less} than
0.05 that the distributions (red and black histograms) are drawn from
the \emph{same} parent distribution.  The constancy of the M/L ratio
and SSFR between $z\simeq 1.5$ and $5$ agrees very well with the
constant slope we find in \figref{fig:mass_abs} and
\figref{fig:mass_sfr} for our sample of LBGs, though with a slightly
larger scatter.

Stellar masses of LBGs at \uvdrops\ are generally well correlated with
UV absolute magnitude and current SFR, as expected for star-forming
galaxies at similar redshifts
\citep[e.g.,][]{elba07,dadd07,sawi12}. These correlations implies very
similar mass assembly and SFH for these galaxies, but the exact nature
of SFHs is still not clearly understood.  \citet{papo11} showed that
the cosmologically averaged SFRs of star-forming galaxies at $3 < z <
8$ --- at constant co-moving number density --- increase smoothly from
$z=8$ to $3$, and the stellar mass growth in these galaxies is
consistent with this derived SFH. The scenario of rising SFH
\citep[see also][]{lee10} is also supported by recent results from the
cosmological hydrodynamic simulations \citep[e.g.,][]{finl11}.  The
models with rising SFHs conflicts with the assumptions that the SFR in
distant galaxies is either constant or decreasing exponentially with
time \citep[e.g.,][]{papo05,shap05,labb10}. Though, we remind the
reader that the models with rising SFHs advocated by \citet{papo11}
and others correspond to a cosmologically averaged SFHs for typical
galaxies, and not individual galaxies, because they could involve
random events that changes their instantaneous SFR. \citet{papo11} and
\citet{lee10} also argue that rising SFHs are most beneficial to
higher redshift ($z\gtrsim 3$) galaxies. We find that for our assumed
SED model parameters, the LBGs between redshift $z\simeq 1.5$ and $5$
--- on average --- have similar SFHs, though the precise nature of
SFHs at all redshift is still under debate, and could also affect the
SFR--stellar mass correlation.

Our analysis demonstrates that the dropout selected galaxies at
\uvdrops\ --- within luminosities probed here --- show similar
correlations between physical parameters (SFR, stellar mass, UV
luminosity) as other star-forming galaxies selected using different
color criteria (e.g., $sBzK$, BX/BM) at $z\simeq 2$. This is
consistent with the \citet{ly11} conclusion that majority
($\sim$80--90\%) of the dropout selected galaxies overlap with other
color selected star-forming galaxies with stellar masses less than
10$^{10}$~M$_{\odot}$.  The stellar mass range for our current sample
is between $\sim$10$^{8}$ and $\sim$10$^{10}$ M$_{\odot}$, with sample
completeness around 10$^{9-9.5}$ M$_{\odot}$.  Significant differences
between the dropout selected sample and other color selected samples
of star-forming galaxies at $z\simeq 2$  exists for massive galaxies
($\gtrsim\,$10$^{10}$ M$_{\odot}$; \citealt{ly11}). Therefore, it is
vital to use uniform selection technique at all redshifts to avoid any
selection biases.  The Lyman break dropout technique is the most
convenient and widely used method to select galaxies at $z\gtrsim 3$,
and we have shown that LBGs at \uvdrops\ selected using this dropout
technique have similar physical properties (within uncertainties) as
LBGs at \bvdrops\ with similar luminosities.  Hence, LBG selection at
$z<3$ is important to understand properties of LBGs and
properly investigate their evolution as a function of redshift.  The
validity of LBG properties over wide luminosity and mass range can be
investigated in detail with the upcoming and future WFC3 UV surveys
such as CANDELS \citep{grog11,koek11} and the WFC3 UV UDF \citep[PI:
H. Teplitz]{rafe12}.


\section{Summary}\label{conclusion}

In this paper, we investigated stellar populations of LBGs at
\uvdrops\ selected using \emph{HST}/WFC3 UVIS filters in the GOODS-S
field. We used deep multi-wavelength observations from the \emph{HST}, VLT,
and \emph{Spitzer} to compare observed SEDs with the spectral synthesis
models to infer physical properties (stellar masses, stellar ages,
SFRs, and dust extinction) of these LBGs. We also compared these LBGs
with their higher redshift (\bvdrops) counterparts with similar
luminosities ($0.1L^*\lesssim L \lesssim 2.5L^*$). Our results can be
summarized as follows:

$\bullet$ We obtain reliable ($\sigma$($z_{sp} -
z_{ph}$/1+$z_{sp}$)$\simeq$0.05) photometric redshifts for dropout
selected LBGs at \uvdrops\ based on 10--13 band SEDs.

$\bullet$ The UV continuum slope $\beta$ for LBGs at \uvdrops\ is
redder ($\beta\simeq\,$--1.6 at $z\simeq 1.6$) compared to their
higher redshift counterparts ($\beta\simeq\,$--2.4 at $z\simeq 8$),
implying higher dust content in these LBGs.

$\bullet$ On average, LBGs at \uvdrops\ are massive, dustier and more
highly star-forming compared to LBGs at \bvdrops, though their median
values are very similar within estimated 1$\sigma$ uncertainties.
This similarity emphasizes the importance of identical Lyman break
selection technique at all redshifts, which selects physically similar
galaxies.

$\bullet$ The stellar mass--absolute UV magnitude relation for LBGs
between $z\simeq 1.5$ and $5$ show linear correlation with a
logarithmic slope of $\sim$0.46, while the SFR--stellar mass relation
show similar correlation with a logarithmic slope of $\sim$0.90. To
properly compare and interpret such relations at higher ($z>3$)
redshift, and to avoid any selection biases due to different selection
techniques, a true Lyman break selection is required at $z\simeq 2$.

$\bullet$ We need larger \emph{HST} UV surveys to cover full range in
luminosity/mass and better understand LBG properties, and their
evolution. Both deeper and wider UV surveys are needed. The wider one
to probe the high mass end, while the deeper one will probe the
sub-$L^*$ population. A large number of \emph{HST} orbits have been used
for dropout selected galaxies at $z>3$, and the lower redshift regime
needs to be explored in a comparable manner.

\acknowledgments

We thank the referee for helpful comments and suggestions that
significantly improved this paper.
This paper is based on Early Release Science observations made by the
WFC3 Scientific Oversight Committee. We are grateful to the Director
of the Space Telescope Science Institute for awarding Director's
Discretionary time for this program. Support for program \#11359 was
provided by NASA through a grant HST-GO-11359.08-A from the Space
Telescope Science Institute, which is operated by the Association of
Universities for Research in Astronomy, Inc., under NASA contract NAS
5-26555. This research was (partially) supported by a grant from the
American Astronomical Society.



\clearpage

\clearpage

\clearpage

\clearpage

\clearpage

\clearpage

\clearpage

\clearpage

\clearpage

\begin{deluxetable}{ccccc}
\tablewidth{0pt}
\tabletypesize{\footnotesize}
\tablecaption{UV Spectral Slopes ($\beta$) \label{tab:beta}}
\tablenum{1}
\tablehead{\colhead{Redshift} &
  \colhead{Number of} &
  \colhead{UV Slope$^a$} & \colhead{1$\sigma$} &
  \colhead{SEM$^b$} \\
 $<z>$ & Galaxies (N) & $\beta$  & scatter & }

\startdata
1.6 ($\pm$0.2) & 47 & --1.59 (--1.67)  & 0.29 (0.27)  & 0.04 (0.04) \\
2.2 ($\pm$0.3) & 126 & --1.71 (--1.81)   & 0.34 (0.37) & 0.03 (0.03) \\
2.6 ($\pm$0.4) & 213 & --1.71 (--1.78)   & 0.47 (0.50) & 0.03 (0.03) \\
3.8$^c$ ($\pm$0.5) & 182 & --1.88 (--1.93)  & 0.45 (0.41) & 0.03 (0.03) \\

\enddata

\tablenotetext{a}{The UV continuum slope $\beta$ is estimated from best-fit SEDs by
  fitting a line between the rest-frame wavelengths 1300--1900~\AA\ (1300--3400~\AA)}
\tablenotetext{b}{The standard error of the mean = (1$\sigma$
  scatter)/$\sqrt{N}$}
\tablenotetext{c}{The comparison sample of \acsb- and
  \acsv-dropouts. Because of small number of \acsv-dropouts, the
  average/median redshift is similar to a \acsb-dropout.}

\end{deluxetable}

\end{document}